\documentstyle[manuscript,aps]{revtex}
\newcommand{\beq}{\begin{equation}}
\newcommand{\eeq}{\end{equation}}
\newcommand{\bea}{\begin{eqnarray}}
\newcommand{\eea}{\end{eqnarray}}
\newcommand{\beqa}{\begin{eqnarray}}
\newcommand{\eeqa}{\end{eqnarray}}

\def \H {{\bf H}}

\def \la {\langle}
\def \ra {\rangle}

\def \T {{\bf T}}
\def \t {|\tau^+_\Delta\ra}
\def \tconj {\la\tau^+_\Delta |}
\def \la {\langle}
\def \ra {\rangle}
\def \f {{\bf f}}

\def \e {\epsilon}
\def \f {f_\epsilon(k)}
\def \go {{_o g}_{t_A}^+(k)}
\def \ge {{_\epsilon} g_{t_A}^+(k)}
\input epsf

\begin{document}
\pagenumbering{arabic}
\title{Time-of-Arrival States}

 \author{ { 
J. Oppenheim$^{(a)}$,\footnote{\it jono@physics.ubc.ca}
B. Reznik$^{(b)}$,\footnote{\it reznik@t6-serv.lanl.gov}
and W. G. Unruh$^{(c)}$}\footnote{\it unruh@physics.ubc.ca \\
}  
{\ } \\
(a) {\it \small   Department of Physics and Astronomy,  University of British
Columbia,
6224 Agricultural Rd. Vancouver, B.C., Canada
V6T1Z1}\\
(b) {\it \small Theoretical Division, T-6, MS B288, 
Los Alamos National Laboratory, Los Alamos, NM, 87545}\\
(c) {\it \small   CIAR Gravity and Cosmology Program, Department of Physics
and Astronomy,  University of British
Columbia,
6224 Agricultural Rd. Vancouver, B.C., Canada
V6T1Z1}
}

\maketitle

\begin{abstract}
Although one can show formally that a time-of-arrival operator cannot
exist, one can modify the low momentum behaviour of the operator slightly 
so that it is self-adjoint.  We show that such a modification
results in the difficulty that the eigenstates are drastically altered. 
In an eigenstate of the modified time-of-arrival operator, the particle,
at the predicted time-of-arrival, is found far away from the point of
arrival with probability $1/2$   
\end{abstract}
\section{Introduction}
In quantum mechanics, observables like position and momentum are
represented by operators at a fixed time $t$.  However, there is no 
operator associated with the time it takes for a particle
to arrive to a fixed location. 
One can construct such a time-of-arrival operator \cite{top},
but its physical meaning is ambiguous \cite{allcock1}
\cite{aharonov}\cite{tmeas}.  
In classical mechanics, one can answer the question, "at what time
does a particle reach the location $x=0$?", but in quantum mechanics,
this question does not appear to have an unambiguous answer.
In \cite{aharonov} we proved formally, that in general 
a time-of-arrival operator cannot exist.  This is because one can prove
that the existence of a time-of-arrival operator implies the
existence of a time operator.  As Pauli \cite{pauli} showed,
one cannot have a time operator if the Hamiltonian of the system
is bounded from above or below.

There has however been renewed interest in time-of-arrival,
following the suggestion by Grot, Rovelli, and Tate, that one can 
modify the time-of-arrival
operator in such away as to make it self-adjoint \cite{rovelli}.  
The idea is that by modifying the operator in
a very small neighbourhood around $k=0$, one can formally construct
a modified time-of-arrival operator which behaves in much the same
way as the unmodified time-of-arrival operator.

In this paper, we examine the behaviour of the modified time-of-arrival
eigenstates, and show that the modification, no matter how small,
radically effects the behaviour of the states.  We find that the
particles in these eigenstates don't arrive with a probability of $1/2$ at
the predicted time-of-arrival.  

In Section II we show why the time-of-arrival operator is not self-adjoint,
and explore the possible modifications that can be made in order to make
it self-adjoint.  We then explore some of the properties of the
modified time-of-arrival states. In Section III we examine normalizable
states which are coherent superpositions of time-of-arrival eigenstates,
and discuss the possibility of localizing these 
states at the location of arrival at the time-of-arrival.  These
results seem to agree with those of Muga and Leavens who have studied 
these states independently \cite{tsups}.
Our central result is contained in Section IV where we show that 
in an eigenstate of the modified time-of-arrival operator, the particle,
at the predicted time-of-arrival, is found far away from the point of
arrival with probability $1/2$  
We also calculate the average energy of the states, in order to relate
them to our proposal \cite{aharonov} that one cannot measure the 
time-of-arrival to an accuracy better than $1/\bar E_k$ where $\bar E_k$ is
the average kinetic energy of the particle.  We finish with some concluding
remarks in Section V.
\section{The Time-of-Arrival Operator \label{top}}
Classically, the position of a free particle is given by
\beq
x(t) = \frac{p_o t}{m} + x_o
\eeq
One can invert this equation to find the time that a particle arrives
to a given location.  From the correspondence principal, one can then
try to define  a time-of-arrival operator $\T$.  The time-of-arrival operator
to the point $x=0$ can be written in the 
$k$ representation as
\beq
\T(k) = 
-im\frac{1}{\sqrt{k}}\frac{d}{dk}\frac{1}{\sqrt{k}}    \, 
\eeq
where $\sqrt{k}=i\sqrt{|k|}$ for $k<0$. It can be verified that the
eigenstates of this operator are given by
\beq
g_{t_A}(k) =  \alpha(k) \frac{1}{\sqrt{2\pi m}}
\sqrt{k}e^{i\frac{t_A k^2}{2m}}   
\eeq
where $\alpha=\left(  \theta(k)+i\theta(-k) \right)$.
These eigenstates however, are not orthogonal.  
\beqa
\langle t_A' | t_A \rangle 
&=&
\frac{1}{\sqrt{2\pi m}}\int_0^\infty dk^2 \, e^{\frac{i}{2m}k^2(t_A-t_A')}
\nonumber\\
&=&
\delta(t_A-t_A') - \frac{i}{\pi(t_A-t_A')} \label{eq:nonortho}.
\eeqa
The reason for this, is that the adjoint of $\T$  
has a different domain of definition than $\T$ itself.  
If $\T$ is defined over all 
square integrable, differentiable functions $v(k)$, then the quantity
\beqa
\la u, \T v \ra - \la \T ^* u, v \ra 
& = &
-i \, m \int dk\left[ 
\frac{\overline{u(k)} }{\sqrt{k}}\frac{d}{dk}\frac{v(k)}{\sqrt{k}} 
+ v(k) \overline{  \frac{1}{\sqrt{k}}\frac{d}{dk}\frac{u(k)}{\sqrt{k}}  }
\right]
\nonumber\\
&=&
i \, m \int_{-\infty}^{0^-} dk\left[ 
\frac{\overline{u(k)} }{\sqrt{|k|}}\frac{d}{dk}\frac{v(k)}{\sqrt{|k|}} 
+ v(k) \overline{  \frac{1}{\sqrt{|k|}}\frac{d}{dk}\frac{u(k)}{\sqrt{|k|}}  }
\right] \,\, - 
\nonumber\\
&&
i \, m \int_{0^+}^{\infty} dk\left[ 
\frac{\overline{u(k)} }{\sqrt{k}}\frac{d}{dk}\frac{v(k)}{\sqrt{k}} 
+ v(k) \overline{  \frac{1}{\sqrt{k}}\frac{d}{dk}\frac{u(k)}{\sqrt{k}}  }
\right]
\nonumber\\
&=&
i\,m\left[
\lim_{k\rightarrow 0^-}
\frac{v(k)\overline{u(k)}}{|k|} + 
\lim_{k\rightarrow 0^+}\frac{v(k)\overline{u(k)}}{|k|} 
\right] 
\label{eq:notadjoint}
\eeqa
will only vanish if $\frac{v(k)\overline{u(k)}}{k}$ is continuous through
$k=0$.  Since $v(k)$ is arbitrary, $\T^*$ is only defined for functions $u(k)$ such that $u(k)/k$ is continuous.  On the other hand, if we change the domain
of definition of $\T$ so that it is defined on functions $v(k)$ such that
$\frac{v(k)}{\sqrt{k}}$ is continuous through $k=0$, then $\T^*$ will only be defined on functions $u(k)$ such that 
$\frac{u(k)}{\sqrt{k}}$ is anti-continuous.  
The domain of definition of $\T$ and $\T^*$ are different,
and thus $\T$ is not self-adjoint.
The problem is not that $\T$ is singular at
$k=0$, but rather that it changes sign discontinuously.    
In some sense, it is like trying to define $-i\, d/dk$ with different 
sign for positive and negative values of $k$.  $-i\, d/dk$ cannot be
defined only on half the real line because it is the generator of 
translations in $k$.  The inability to define a self-adjoint operator
$\T$ is directly related to the fact that one cannot construct an operator
which is conjugate to the Hamiltonian if $\H$ is bounded from below \cite{aharonov}.

One might therefore try to modify the time-of-arrival operator, in such
a way as to make it self-adjoint \cite{rovelli}.
Consider the operator
\beq
\T_\e (k) = 
-im\sqrt{\f}\frac{1}{dk}\sqrt{\f}     \,\, 
\eeq
where $\f$ is some smooth function.  
Since $u(k)$ and $v(k)$ could diverge at the origin
at a rate approaching $1/\sqrt{k}$ and still remain square-integrable, 
if $\f$ goes to zero at least as fast as
$k$, then $\T_\e$ will be self-adjoint and defined over all square
integrable functions.  
It can then be verified that it has a degenerate set of  
eigenstates   $|t_A,+\ra$ for $k>0$ and $|t_A,-\ra$ for $k<0$,
given by
\beq
g_{t_A}^\pm(k) = \theta(\pm k)\frac{1}{\sqrt{2\pi m}} \frac{1}{\sqrt{\f}}
e^{\frac{it_A}{m}\int_\e^k f_\e(k')dk'} \,\,\,\,   
\eeq
Grot, Rovelli, and Tate \cite{rovelli} choose to work with the states
given by 
\beq
 \f = \left\{ \begin{array}{ll}
\frac{k}{\e^2} & \,\,\,\, |k| < \e \\
\frac{1}{k} & \,\,\,\, |k| > \e 
\end{array}
\right. 
\eeq
When $\e\rightarrow 0$, it is believed that the modification will not 
effect measurements of time-of-arrival if the state does not have
support around $k=0$ \cite{rovelli}. 

As mentioned, if the domain of definition of $\T_\e$ is smooth, square-integrable functions, than any $\f$ which went to zero slower 
than this choice
would not be sufficient.  Also, as we will show in the Section IV,
any function which goes to zero faster than $k$ will have the problem
that a particle in an eigenstate of the modified time-of-arrival operator
will have a greater chance of not arriving at the predicted time.  
We therefore will also choose to work with this function.
Explicitly,
we see that the eigenfunctions are now given by
\beq
g_{t_A}^\pm(k) \equiv {_o g}_{t_A}^\pm(k)
+ {_\epsilon} g_{t_A}^\pm(k)
\eeq
where for example
\beq
\ge = \left\{ \begin{array}{ll}
\frac{1}{\sqrt{2\pi m}} \frac{1}{\sqrt{k}}
e^{\frac{it_A}{m} \ln{k/\e}}  & \,\,\,\, |k| < \e \\
0  & \,\,\,\, |k| > \e 
\end{array}
\right. 
\eeq

\beq
 \go = \left\{ \begin{array}{ll}
0 & \,\,\,\, |k| < \e \\
\frac{1}{\sqrt{2\pi m}} {\sqrt{k}}
e^{\frac{i t_A}{2m} (k^2-\e^2)} & \,\,\,\, |k| > \e 
\end{array}
\right. 
\eeq
In the limit $\e\rightarrow 0$, 
$\go$ behaves in a manner which one might associate with a 
time-of-arrival state, while $\ge$ is due to the modification of $\T$.
Grot, Tate, and Rovelli show that these eigenstates are orthogonal by writing
them in the coordinates
\beq
z^\pm=\int_{\pm \e}^{k} \frac{dk'}{f_\e(k')} \, \, .
\eeq
These coordinate go from $-\infty$ to $\infty$.  We can now see that 
these modified eigenstates are orthogonal:
\beqa
\la t_A',\pm | t_A,\pm\ra 
&=& 
\int_{-\infty}^\infty dz^\pm 
e^{i(t_A-t_A')\frac{z^\pm}{m}}
\nonumber\\
&=&
\delta(t_A-t_A') \,\, .
\label{eq:ortho}
\eeqa
The states $|t_A, + \ra$ and $|t_A, -\ra$ can also be shown to be orthogonal.

When these states are examined in the x-representation,
one can see that at the time-of-arrival, the functions $\go$ are 
not delta functions
$\delta(x)$ but are proportional to $x^{-3/2}$; 
it has support over all $x$ \cite{aharonov}.
However, although the state has long tails out to infinity, the quantity
$\int dx' |x'^{-3/2}|^2 \sim x^{-2}$ 
goes to zero as $x\rightarrow\infty$.  Furthermore, the modulus squared
of the eigenstates diverges when integrated around the point of arrival $x=0$.  As a result, the normalized 
state will be localized at the point-of-arrival at the time-of-arrival.
In Section  III we show that this is indeed so.  On the other
hand, the Fourier-transform of the state $\ge$ at the 
time-of-arrival is given by
\beq
{_\e{\tilde g}^+(x)_{t_A}}
=
\frac{\e}{\sqrt{2\pi m}}
\int_0^\epsilon \frac{dk}{\sqrt{k}} 
e^{ikx}e^{-it_A\frac{k^2}{2m}}
e^{\frac{i\e^2 t_A}{m}
\ln{\frac{k}{\e}}}
\eeq
Because $\T_\e$ is no longer the generator of energy translations for $|k|<\e$,
$\ge$ is not time-translation invariant.
For the $t_A=0$ state, this can be integrated to give
\beq
{_\e{\tilde g}^+(x)_{t_A}}
=
\frac{\e}{\sqrt{2xim}}\Phi(\sqrt{i\e x}) \label{eq:gex}
\eeq
where $\Phi$ is the probability integral.  For large $x$, 
$_\e{\tilde g}_{t_A}^+ (x)$
goes as $\frac{1}{\sqrt{x}}$ and the quantity  
$\int dx' |x'^{-3/2}|^2 \sim \ln{x}$
diverges as $x\rightarrow\infty$.  For small $x$, 
$_\e{\tilde g}_{t_A}^+ (x)$ is proportional to $  e^{-i\e x}$. 
Its modulus squared
vanishes when integrated around a small neighbourhood of $x=0$.  
$\ge$ then, is not localized around the point of arrival, at the 
time-of-arrival.  This will also be verified in Section III where we 
examine the normalizable states.  Although $\ge$ is not localized around
the point of arrival at the time of arrival, one might hope that this part of
the state does not contribute significantly in time-of-arrival measurements
when $\e \rightarrow 0$.

\section{Normalized Time-of-Arrival States \label{normstates}}
Since the time-of-arrival states are not normalizable, we will examine
the properties of states
$|\tau_\Delta \ra$ which are narrow superpositions of the time-of-arrival eigenstates.  These states are normalizable, although they are no longer orthogonal to each other
\footnote{These coherent states form a positive operator valued measure
(POVM).  While there are no self-adjoint time-of-arrival operators,
time-of-arrival may be represented by POVMs
\cite{povm}.}
.  By decreasing $\Delta$, the spread in arrival-times, $|\tau_\Delta \ra$ must be as localized as one wishes around the point
of arrival, at the time-of-arrival.  They must also have the feature  
that at times other than the
time-of-arrival, one can make the probability that the particle is found
at the point of arrival vanish as $\Delta$ goes to zero.

We can now consider coherent states of these eigenstates 
\beq
|\tau_\Delta^\pm \ra = N \int dt_A |t_A,\pm\ra e^{-\frac{(t_A-\tau)^2}{\Delta^2}}.
\eeq
where $N$ is a normalization constant and is
given by $N=\frac{(2\pi^3)^{-1/4}}{\sqrt{\Delta}} $.
The spread $dt_A$ in arrival times is of order $\Delta$.  

We now examine what the state $\tau(x,t)^+=\la x\t$ 
looks like at the point of arrival 
as a function of time.   In what follows, we will work with the state
centered around $\tau =0$ for simplicity.  This will not affect any of our conclusions. 
$\tau^+(x,t)$ is given by
\beqa
\tau^+(x,t)
&=&
N \int   
\la x |e^{\frac{-i{\bf p}^2 t}{2m}}|t_A,+
\ra e^{-\frac{t_A^2}{\Delta^2}}dt_A
\nonumber\\
&=& 
N \int_0^\e  e^{-\frac{t_A^2}{\Delta^2}} 
e^{\frac{-ik^2}{2m}t} e^{ikx}  \ge dt_A \, dk 
\,\,\,\,\, +
N \int_\e^\infty e^{-\frac{t_A^2}{\Delta^2}} 
e^{\frac{-ik^2}{2m}t} e^{ikx}\go dt_A \, dk
\nonumber\\
&\equiv&
{_\e\tau}^+(x,t) \, \, +
{_o\tau}^+(x,t) 
\eeqa
As argued in the previous section, the second term 
should act like a time-of-arrival state.  
The first term is due to the modification of $\T$ and has nothing
to do with the time of arrival.  We will first show that the second
term can indeed be localised at the point-of-arrival $x=0$ at the time
of arrival $t=t_A$.  We will do this by expanding it around $x=0$ in
a Taylor series.  After taking the limit $\e\rightarrow 0$,
it's n'th derivative at $x=0$ is given by  
\beqa
\frac{d^n}{dx^n}   {_o\tau}^+(x,t)|_{x=0}
&=&
\frac{N}{\sqrt{2\pi m}} 
\int e^{-\frac{t_A^2}{\Delta^2}}\theta(k) \sqrt{k} (i k)^n
e^{\frac{ik^2}{2m}(t_A-t)} dt_A \, dk
\nonumber\\
&=&
\frac{N\Delta}{\sqrt{2m}}i^n \int^\infty_0 
e^{\frac{-k^4\Delta^2}{16m^2}}
e^{\frac{-ik^2 t}{2m}}
k^{\frac{1}{2}+n} dk
\nonumber\\
&=&
\frac{i^n}{2} \frac{N \Delta}{\sqrt{2m}}
\int_0^\infty 
e^{\frac{-{\tilde k}^2\Delta^2}{16m^2}}
e^{\frac{-i{\tilde k} t}{2m}}
{\tilde k}^{\frac{1}{4}+\frac{n}{2}} d{\tilde k}
\nonumber\\
&=&
\frac{2^{-\frac{1}{8}+\frac{3n}{4}} i^n}{\pi^{\frac{3}{4} } }
\Gamma(\frac{3}{4} + \frac{n}{2})
(\frac{m}{\Delta})^{\frac{1}{4}+\frac{n}{2}}
e^{-\frac{t^2}{2\Delta^2}} 
D_{-\frac{3}{4}-\frac{n}{2}}(\frac{it\sqrt{2}}{\Delta}) \label{eq:unmodt}
\eeqa
where $D_p(z)$ are the parabolic-cylinder functions.
For any finite $t$, we can choose $\Delta$ small enough so that
the argument of $D_p (z)$ is large, and can be expanded as
\beq
D_{p}(z) \simeq e^{-\frac{z^2}{4}}z^p(1-\frac{p(p-1)}{2z^2} 
+ \cdot \cdot \cdot)
\eeq
so that $\frac{d^n}{dx^n}{_o\tau}^+(x,t)|_{x=0}$ behaves as
\beq
\frac{d^n}{dx^n}{_o \tau}^+(x,t)|_{x=0} 
\simeq a_n \frac{{\sqrt{\Delta} m^{\frac{1}{4}+\frac{n}{2}  }}}
{t^{\frac{3}{4}+\frac{n}{2}}} \, .
\eeq
where $a_n$ is a numerical constant given by
\beq
a_n = i^{-\frac{3}{4} +\frac{n}{2}}2^{\frac{n}{2} -1}\pi^{-\frac{3}{4}} 
\Gamma(\frac{3}{4}+\frac{n}{2})  
\eeq 
We can now write $_o\tau^+(0,t)$ as a Taylor expansion around $x=0$
\beq
_o\tau^+(x,t) \simeq \sqrt{\Delta} (\frac{m}{t^3})^\frac{1}{4} 
\sum_{n=0}^{\infty} 
a_n (\sqrt{\frac{m}{t}} x)^n 
\eeq
We can now see that
for any finite $t$ the amplitude for
finding the particle around $x=0$ goes to zero as $\Delta$ goes to zero.  
The probability of being found at the point of arrival at a time other 
than the time-of-arrival can be made arbitrarily small.
On the other hand,
at the time-of-arrival $t=0$, we will now show that the particle can
be as localized as one wishes around $x=0$.

From (\ref{eq:unmodt}), we expand $_o\tau^+(x,0)$ as a Taylor series
\beq
_o\tau^+(x,0) =(\frac{m}{\Delta})^\frac{1}{4} 
\sum_{n=0}^{\infty} 
b_n (\sqrt{\frac{m}{\Delta}}x)^n
\eeq
where
\beqa
b_n & = &
i^n 2^{-\frac{5}{8}+\frac{3n}{4}} \pi^{-\frac{3}{4}}
\Gamma(\frac{3}{4}+\frac{n}{2}) D_{-\frac{3}{4} -\frac{n}{2}}(0)
\nonumber\\
&=&
i^n 2^{n-\frac{5}{4}} \pi^{-\frac{3}{4}}
\Gamma(\frac{3}{8}+\frac{n}{4})
\eeqa
We see than that $_o\tau^+(x,0)$ is a function of 
$\sqrt{\frac{m}{\Delta}}x$ (with a constant of  
$(\frac{m}{\Delta})^{1/4}$ out front). 
As a result, the probability of finding the particle in a neighbourhood
$\delta$ of $x$ is given by
\beq
\int_{-\delta}^\delta |{_o\tau}^+ (\sqrt{\frac{m}{\Delta}}x,0)|^2 dx 
=
\sqrt{\frac{\Delta}{m}}
\int_{-\delta\sqrt{\frac{m}{\Delta}}}^{\delta\sqrt{\frac{m}{\Delta}}} 
|{_o\tau}^+ (u,0)|^2 du .
\eeq
Since  $|{_o\tau}^+(u,0)|^2$ is proportional to $\sqrt{\frac{m}{\Delta}}$,
and is square integrable, we see that for any $\delta$, one need only
make $\Delta$ small enough, in order to localize the entire particle
in the region of integration. 
The state $_o\tau^+(x,t)$ is localized in a neighbourhood $\delta$
around the point-of-arrival at the time-of-arrival as $\Delta \rightarrow 0$.  
The state is localized in a region $\delta$ of order 
$\sqrt{\frac{\Delta}{m}}$.
This is what one would expect from physical grounds, since we have
\beqa
dx &\sim& dt_A\frac{\la k \ra}{m} \nonumber\\
&\sim& \sqrt{\frac{\Delta}{m}} \,\, .
\eeqa
($\la k \ra$ is calculated in the following section and is proportional
to $\sqrt{m/\Delta}$).
The probability distribution of $_o\tau^+(x,t)$ at $t=\tau$ is shown in Figure 1.   This behaviour of 
$_o\tau^+(x,t)$ as a function of time appears to agree with the results of 
Muga and Leavens, who have studied these coherent
states independently \cite{tsups}.

The state ${_\e \tau^+(x,0)}$ is not found near the origin at
$t=t_A=0$.  We find
%

\beqa
{_\e\tau}^+(x,0)
&=&
N \frac{\e}{\sqrt{2\pi m}}
\int_{-\infty}^\infty \int_0^\e e^{-\frac{t_A^2}{\Delta^2}} 
\frac{1}{\sqrt{k}} 
e^{\frac{i\e^2t_A}{m}\ln{\frac{k}{\e}}} e^{ikx}dk \, dt_A
\nonumber\\
&=&  
N \frac{\e^{3/2}}{\sqrt{2\pi m}}
\int_{-\infty}^\infty \int_0^1 
e^{-\frac{t_A^2}{\Delta^2}} k^{\frac{i\e^2 t_A}{m} -\frac{1}{2}}  
e^{ik\e x}dk \, dt_A
\nonumber\\
&=&  
N \frac{\e^{3/2}}{\sqrt{2\pi m}}
\int_{-\infty}^\infty  
e^{-\frac{t_A^2}{\Delta^2}} 
\gamma(\frac{i\e^2 t_A}{m} +\frac{1}{2},-i\e x)  
(-i\e x)^{-\frac{1}{2}-\frac{i\e^2 t_a }{m}} dt_A. \label{eq:modstat}
\eeqa
If $i\e x$ is not large, we can use the fact that for $\Delta$ and $\e$ very small, $i\e^2 t_A/m \ll 1/2$ so that
we have
\beqa
{_\e\tau}^+(x,0)
&\simeq&
N \frac{\e^{3/2}}{\sqrt{2\pi m}}  
\frac{\gamma(\frac{1}{2},-i\e x)}{\sqrt{-i\e x}}
\int_{-\infty}^\infty  
e^{-\frac{t_A^2}{\Delta^2}} dt_A
\nonumber\\
&=&
(2\pi)^{-\frac{1}{4}}\sqrt{ \frac{\e^3\Delta}{2m} }  
\frac{\Phi(\sqrt{-i\e x})}{\sqrt{-i\e x}} \, . 
\eeqa
Note the similarity between this state (the form above is not valid
for large $x$), and that of the modified part of the eigenstate (\ref{eq:gex}). 
We are interested in the case where $\frac{\e^2 \Delta}{m}$ goes to zero, 
in which case $_\e\tau^+(x,0)$ vanishes near the origin.  For large $\e x$,
it goes as  $\sqrt{ \frac{\e^2\Delta}{x m} } $.
From (\ref{eq:modstat}) we can also see that if 
$\e x>e^{\frac{m}{e^2\Delta}}$ then the last factor in the integrand 
oscillates rapidly
and the integral falls rapidly for larger $x$.  
Thus, as we make $\frac{\e^2 \Delta}{m}$
smaller, the value of the modulus squared decrease around $x=0$,
but the tails, which extend out to $e^{\frac{m}{e^2\Delta}}/\e$, 
get longer. $\int^x |_\e\tau^+(x,0)|^2$ goes as $\frac{\e^2 \Delta}{m} \ln{x}$
up to $\e x \sim e^\frac{m}{\e^2 \Delta}$.

As $\frac{\e^2 \Delta}{m}\rightarrow 0$, the particle
is always found in the far-away tail.
The state $\ge$ is not found
near the point of arrival at the time-of-arrival.  It's probability
distribution at $t=t_A=0$ is shown in Figure 2.
%
%
%
%
%
\section{Contribution to the Norm due to Modification of $\T$}
We now show that the modified part of $\t$ contains half the norm,
no matter how small $\e$ is made.
The norm of the state $\t$ can be written as  
\beqa
\int |\la x | \tau^+_\Delta \ra |^2 dx
&=&
N^2\int |\la x |k\ra  e^{-\frac{t_A^2}{\Delta^2}} \go
dt_A dk |^2 dx \, \, \, +
N^2\int |\la x |k\ra  e^{-\frac{t_A^2}{\Delta^2}} \ge
dt_A dk |^2 dx
\nonumber\\
&\equiv& N_o^2 + N_\e^2
\eeqa
where $N_o^2$ is the norm of the unmodified part of the time-of-arrival
state, and $N^2_\e$ is the norm of the modified part.  The first term
can be integrated to give
\beqa
N_o^2
&=& 
\frac{N^2}{2\pi m} \int dt_A dt_A' dk dk' dx 
e^{\frac{-t_A^2-t_A'^2}{\Delta^2}} e^{\frac{i(k'^2 t_A' -k^2 t_A)}{2m}} 
e^{ix(k-k')} \theta(k)\theta(k')\sqrt{k}\sqrt{k'} \nonumber
\eeqa
where without loss of generality, we are looking at the state centered around
$\tau=0$ at $t=0$.  Since the integral over $x$ gives the delta 
function $\delta(k-k')$,
we find \
\beqa
N_o^2 
&=& 
\frac{N^2}{m} \int e^{\frac{-t_A^2-t_A'^2}{\Delta^2}} 
e^{\frac{ik^2 }{2m}(t_A' - t_A)} \theta  (k) k dt_A \, dt_A'\, dk
\nonumber\\
&=& 
\frac {N^2\Delta^2 \pi}{m} \int^\infty_0 dk \,  k e^{\frac{-k^4\Delta^2}{8m^2}} 
\nonumber\\
&=& 
\frac {N^2\Delta^2 \pi}{4m} \int^\infty_0 \frac{du}{\sqrt{u}} e^{\frac{-\Delta^2}{8m^2}u} 
\nonumber\\
&=&
\frac{1}{2} \,\, .
\eeqa

The unmodified piece contains only half the norm.  The rest is found
in the modified piece. 
\beqa
N_\e^2
&=& 
\frac{N^2}{2\pi m} \int_0^\e dk dk \int dt_A dt_A'  dx 
e^{\frac{-t_A^2-t_A'^2}{\Delta^2}} e^{\frac{i}{m}(t_A'\ln{\frac{k'}{\e}}
- t_A\ln{\frac{k}{\e}})} 
e^{ix(k-k')} \frac{\e^2}{\sqrt{kk'}} 
\nonumber\\
&=&
\frac{N^2}{ m} \int_0^\e dk  \int dt_A dt_A'  
e^{\frac{-t_A^2-t_A'^2}{\Delta^2}} e^{i\ln{\frac{k}{\e}}\frac{t_A'- t_A}{m}} 
\frac{\e^2}{k} 
\nonumber\\
&=&
\frac{N^2\Delta^2\pi}{ m} \int_0^\e dk    
e^{\frac{-\e^4\Delta^2\ln^2{k/\e}}{2m^2}} 
\frac{\e^2}{k} 
\nonumber\\
&=&
\frac{N^2\e^2\Delta^2\pi}{ m} \int_0^\infty du  \,\,  
e^{\frac{-\e^4\Delta^2}{2m^2}u^2} 
\nonumber\\
&=&
\frac{1}{2} \label{eq:modnorm}
\eeqa

The norm of the modified piece makes up half the norm of the total
time-of-arrival state.  
The reason for this can be seen by examining eqns (\ref{eq:nonortho}) and 
(\ref{eq:ortho}).  The term $\go$ by itself gives
\beq
\int dk {_o g}^+_{t_A}(k) {_o g}^+_{t_A}(k)
=
\frac{1}{2}\delta(t_A-t_A')-\frac{i}{2\pi(t_a-t_A')}
\eeq
as $\e\rightarrow 0$.  The term which contributes another
$\frac{1}{2}\delta(t_A-t_A')$ and cancels the 
principal value $\frac{i}{2\pi(t_a-t_A')}$ term is the modified piece $\ge$.  
Essentially, the modification involves expanding the region
$0<k<\e$ into the entire negative k-axis.
No matter how small we make $\e$, we cannot
avoid the fact that the modified part contributes substantially to
the behaviour of the state.  As a result, if one makes a measurement
of the time-of-arrival, then one finds that half the time, the particle
is not found at the point of arrival at the predicted time-of-arrival.
Modified time of arrival states do not always arrive on time. 

From (\ref{eq:modnorm}), one can also see that if $\f$ goes to zero
faster than $k$, then $N_\e$ will diverge as $\Delta$ or $\e$ go to zero.  
If $\f=k^{1+\delta}$, then we find
\beq
N_\e = 
\frac{1}{2} e^{\frac{\delta^2 m^2}{2\e^4 \Delta^2}}
\left[ 1- \Phi(\frac{-\delta\e^2\Delta\sqrt{2}}{m})\right]
\eeq
As $\e$ or $\Delta$ go to zero, $N_\e$ diverges.  
 
It is also of interest to calculate the 
average value of the kinetic energy for 
these states, since in \cite{aharonov} we found that if one wants to
measure the time-of-arrival with a clock, then the accuracy of the
clock cannot be greater than $1/\bar E_k$.  
In calculating the average energy, the modified piece will not matter
since $k^2$ goes to zero at $k=0$ faster than $\frac{1}{\sqrt{k}}$
diverges.  We find
\beqa
\tconj \H_k \t
&=&   
\int dk \frac{k^2}{2m} \tconj k\ra\la k \t
\nonumber\\
&=&
\frac{N^2}{\pi(2m)^2}\int_0^\infty k^3  e^{\frac{i(t_A-t_A')k^2}{2m}}e^{-\frac{t_A^2+t_A'^2}{\Delta^2}}                                                                                                                                                                                                                       dt_A \, dt_A'\, dk
\nonumber\\
&=&
(\frac{N\Delta}{2m})^2 
\int_0^\infty e^{\frac{-k^4\Delta^2}{8m^2}} k^3  dk
\nonumber\\
&=&   
\frac{2}{\Delta\sqrt{2\pi^3}}
\eeqa
We see therefore, that the kinematic spread in arrival times of these
states is proportional to $1/ \bar E_k$.  Since the probability of 
triggering the model clocks discussed in \cite{aharonov} 
decays as  $\sqrt{E_k \delta t_A}$, where $\delta t_A$ is the accuracy
of the clock, we find that the states $\t$ will not always 
trigger a clock whose accuracy is $\delta t_A = \Delta$.
\section{Conclusion}

We have seen that if one modifies the time-of-arrival operator so as to
make it self-adjoint, then its eigenstates no longer behave as one
expects time-of-arrival states to behave.  Half the time, a particle
which is in a time-of-arrival state will not arrive at the predicted
time-of-arrival.  The modification also results in the fact that the
states are no longer time-translation invariant.

For wavefunctions which don't have support at $k=0$, measurements
can be carried out in such a way that the modification
will not effect the results of the measurement \cite{aharonov}.  
Nonetheless, after the measurement, the particle will not arrive
on time with a probability of $1/2$.  One cannot use $\T_\e$ to prepare
a system in a state which arrives at a certain time.

Previously, we have argued that time-of-arrival measurements should
be thought of as continuous measurement processes, and that there
is an inherent inaccuracy in time-of-arrival measurements, given by
$\delta t_A > 1/\bar E_k$ \cite{aharonov}\cite{traversal}.  This current paper
supports the claim that the time-of-arrival is not a well defined
observable in quantum mechanics.
\\
\\
\\
\\
\\ 
{\bf Acknowledgments}
J.O. and W.G.U. would like to thank Bob Wald for valuable discussion. 
W.G.U. thanks the CIAR and NSERC for support during the completion of
this work.  J.O. also thanks NSERC for their support.   
\vfill \eject

\begin{figure}
\hspace {3.3cm}
\epsfysize=2.0in
\epsfbox{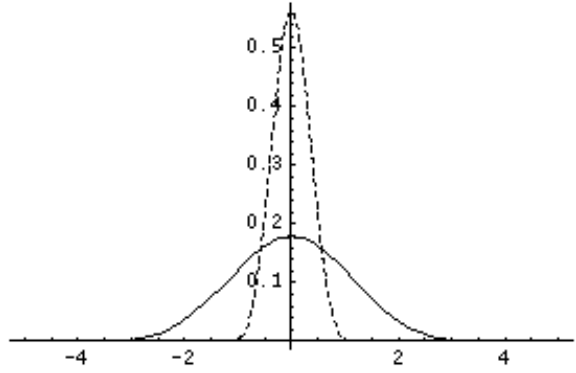}
\caption[]{ 
$|{_o\tau}^+(x,\tau)|^2$ vs. $x$net,
with $\Delta=m$ (solid line), and $\Delta=\frac{m}{10}$ (dashed line).
As $\Delta$ gets smaller, the probability function gets more and more
peaked around the origin.
}
\end{figure}

\begin{figure}
\vspace {3cm}
\hspace {3.3cm}
\epsfysize=2.0in
\epsfbox{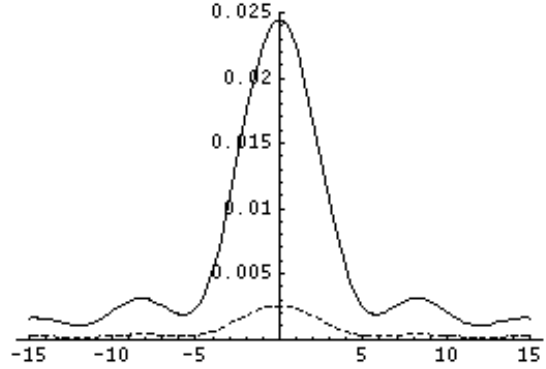}
\caption[]{ 
$\frac{1}{\e}|{_\e\tau}^+(x,\tau)|^2$ vs. $\e x$,
with $\Delta\e^2=\frac{m}{10}$ (solid line) and 
$\Delta\e^2=\frac{m}{100}$ (dashed line).  As $\Delta$ or $\e$ gets
smaller, the probability function drops near the origin, and grows
longer tails which are exponentially far away.
}
\end{figure}


%
%

\end{document}